\documentclass[sigconf]{acmart}

\usepackage{amscd,amsfonts,amsbsy,rotating}
\usepackage{balance}
\usepackage{graphicx}
\usepackage{epsfig,epstopdf}
\usepackage{subfigure}
\usepackage{multirow}
\usepackage{booktabs}
\usepackage{color,xcolor}
\usepackage{url}
\usepackage{latexsym,bm}
\usepackage{enumitem,balance,mathtools}
\usepackage{wrapfig}
\usepackage{euscript}
\usepackage{algorithm}
\usepackage{algorithmic}
\usepackage{ifpdf}
\usepackage{diagbox}
\usepackage{caption}
\usepackage{makecell}
\usepackage{subfigure}
\usepackage{bm}
\usepackage{sidecap}

\usepackage{array}
\newcolumntype{N}{@{}m{0pt}@{}}

\usepackage{lipsum}
\newcommand\blfootnote[1]{%
  \begingroup
  \renewcommand\thefootnote{}\footnote{#1}%
  \addtocounter{footnote}{-1}%
  \endgroup
}

\theoremstyle{plain}

\newtheorem*{definition*}{Definition}

\newcommand{\minisection}[1]{\vspace{5pt}\noindent\textbf{#1.}}

\AtBeginDocument{%
  \providecommand\BibTeX{{%
    \normalfont B\kern-0.5em{\scshape i\kern-0.25em b}\kern-0.8em\TeX}}}

\acmSubmissionID{lp0949}

\copyrightyear{2019}
\acmYear{2019}
\setcopyright{acmcopyright}
\acmConference[CIKM '19] {The 28th ACM International Conference on Information and Knowledge Management}{November 3--7, 2019}{Beijing, China}
\acmBooktitle{The 28th ACM International Conference on Information and Knowledge Management (CIKM'19), November 3--7, 2019, Beijing, China}
\acmPrice{15.00}
\acmDOI{10.1145/3357384.3357978}
\acmISBN{978-1-4503-6976-3/19/11}

\begin{document}
\fancyhead{}
\title{CoRide: Joint Order Dispatching and Fleet Management \\ for Multi-Scale Ride-Hailing Platforms}

\author{Jiarui Jin$^{1}$, Ming Zhou$^1$, Weinan Zhang$^1$, Minne Li$^2$, Zilong Guo$^1$, Zhiwei Qin$^3$, Yan Jiao$^3$, Xiaocheng Tang$^3$, Chenxi Wang$^3$, Jun Wang$^2$, Guobin Wu$^4$, Jieping Ye$^3$}
\affiliation{$^1$Shanghai Jiao Tong University, $^2$University College London, $^3$DiDi AI Labs, $^4$DiDi Research}
\email{{jinjiarui97, mingak, wnzhang, gzlong}@sjtu.edu.cn, {minne.li, jun.wang}@cs.ucl.ac.uk, {qinzhiwei, yanjiao, tangxiaocheng, wangchenxi, wuguobin, yejieping}@didiglobal.com}

\renewcommand{\shortauthors}{J. Jin, et al.}

\begin{abstract}
How to optimally dispatch orders to vehicles and how to trade off between immediate and future returns are fundamental questions for a typical ride-hailing platform.
We model ride-hailing as a large-scale parallel ranking problem and study the joint decision-making task of order dispatching and fleet management in online ride-hailing platforms.
This task brings unique challenges in the following four aspects. 
First, to facilitate a huge number of vehicles to act and learn efficiently and robustly, we treat each region cell as an agent and build a multi-agent reinforcement learning framework. 
Second, to coordinate the agents from different regions to achieve long-term benefits, we leverage the geographical hierarchy of the region grids to perform hierarchical reinforcement learning.
Third, to deal with the heterogeneous and variant action space for joint order dispatching and fleet management, we design the action as the ranking weight vector to rank and select the specific order or the fleet management destination in a unified formulation.
Fourth, to achieve the multi-scale ride-hailing platform, we conduct the decision-making process in a hierarchical way where a multi-head attention mechanism is utilized to incorporate the impacts of neighbor agents and capture the key agent in each scale.
The whole novel framework is named as \emph{CoRide}.
Extensive experiments based on multiple cities real-world data as well as analytic synthetic data demonstrate that \emph{CoRide} provides superior performance in terms of platform revenue and user experience in the task of city-wide hybrid order dispatching and fleet management over strong baselines.

\end{abstract}



\settopmatter{printacmref=false, printfolios=false}


\maketitle

{\fontsize{8pt}{8pt} \selectfont
	\textbf{ACM Reference Format:}\\
	Jiarui Jin, Ming Zhou, Weinan Zhang, Minne Li, Zilong Guo, Zhiwei Qin, Yan Jiao, Xiaocheng Tang, Chenxi Wang, Jun Wang, Guobin Wu and Jieping Ye. 2019. CoRide: Joint Order Dispatching and Fleet Management for Multi-Scale Ride-Hailing Platforms. In \textit{Proceedings of the 28th ACM International Conference on Information and Knowledge Management (CIKM '19), November 3--7, 2019, Beijing, China.} ACM, New York, NY, USA, 10 pages.
	https://doi.org/10.1145/3357384.3357978}

\section{Introduction}
Online ride-hailing platforms such as Uber and Didi Chuxing have substantially transformed our lives by sharing and reallocating transportation resources to highly promote transportation efficiency.
In a general view, there are two major decision-making tasks for such ride-hailing platforms, namely 
(i) order dispatching, i.e., to match the orders and vehicles in real time to directly deliver the service to the users \cite{zou2013novel,zhang2017taxi,seow2010collaborative}, and 
(ii) fleet management, i.e., to reposition the vehicles to certain areas in advance to prepare for the later order dispatching \cite{lin2018efficient,oda2018distributed,simao2009approximate}.\par

\begin{figure}[t]
\centering
\includegraphics[width=0.85\columnwidth]{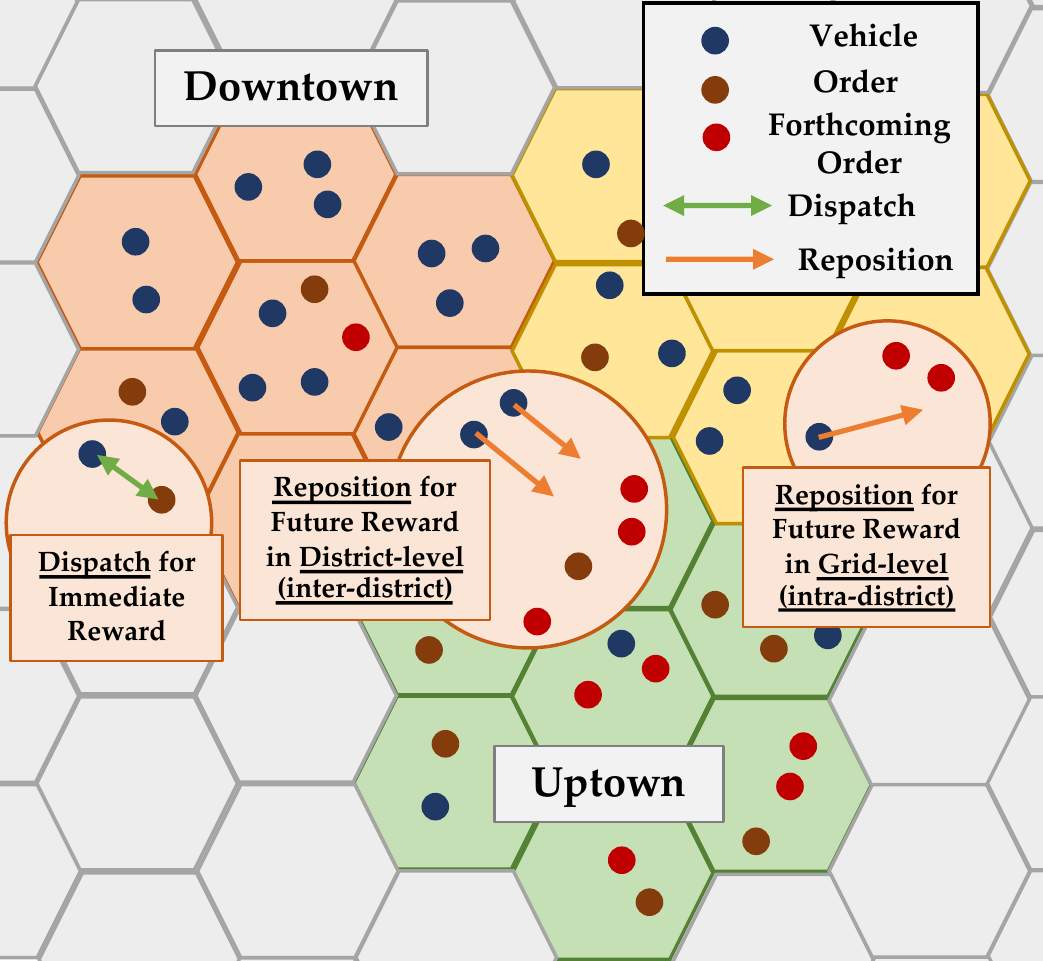}
\vspace{-3mm}
\caption {Ride-hailing task in thermodynamics view.}
\label{fig:thermodynamics}
\vspace{-3mm}
\end{figure}

Apparently, the decision-making of matching an order-vehicle pair or repositioning a vehicle to an area needs accounting for the future situation of the vehicle's location and the orders nearby. 
Thus, much of work has modeled order dispatching and fleet management as a sequential decision-making problem and solved it with reinforcement learning (RL) \cite{wang2018deep,xu2018large,lin2018efficient,tang2018deep}.
Most of the previous work deals with either order dispatching or fleet management without regarding the high correlation of these two tasks, especially for large-scale ride-hailing platforms in large cities, which leads to sub-optimal performance.
In order to achieve near-optimal performance, inspired by thermodynamics, we simulate the whole ride-hailing platform as dispatch (order dispatching) and reposition (fleet management). 
As illustrated in Figure~\ref{fig:thermodynamics},  we resemble vehicle and order as different molecules and aim at building up the system stability via reducing their number by dispatch and reposition. 
To address this complex criterion, we provide two novel views: (i) interconnecting order dispatching and fleet management, and (ii) joint considering intra-district (grid-level) and inter-district (district-level) allocation.
With such a practical motivation, we focus on modeling joint order dispatching and fleet management with multi-scale decision-making system.  
There are several significant technical challenges to learn highly efficient allocation policy for the real-time ride-hailing platform:\par

\minisection{Large-scale Agents}
A fundamental question in any ride-hailing system is how to deal with a large number of orders and vehicles.
One alternative is to model each available vehicle as an agent \cite{oda2018distributed,xu2018large,wei2018look}. 
However, such setting needs to maintain thousands of agents interacting with the environment, which brings a huge computational cost. 
Instead, we utilize the region grid world (as will be further discussed in Figure~\ref{fig:grid}), which regards each region as an agent, and naturally model ride-hailing system in a hierarchical learning setting. 
This formulation allows decentralized learning and control with distributed implementation.\par

\minisection{Immediate \& Future Rewards}
A key challenge in seeking an optimal control policy is to find a trade-off between immediate and future rewards in terms of accumulated driver income (ADI). 
Greedily matching vehicles with long-distance orders can receive high immediate gain at a single order dispatching stage, but it would harm order response rate (ORR) and future revenue especially during rush hour because of its long drive time and unpopular destination. 
Recent attempts \cite{xu2018large,wei2018look,oda2018distributed} deployed RL to combine instant order reward from online planning with future state-value as the final matching value. 
However, the coordination between different regions is still far from optimal.
Inspired by hierarchical RL \cite{vezhnevets2017feudal}, we introduce the geographical hierarchical structure of region agents. 
We treat large district as \emph{manager} agent and small grid as \emph{worker} agent, respectively. 
The \emph{manager} operates at a lower spatial and temporal dimension and sets abstract goals which are conveyed to its \emph{worker}s. 
The \emph{worker} takes specific actions and interacts with environment coordinated with \emph{manager}-level goal and \emph{worker}-level message. 
This decoupled structure facilitates very long timescale credit assignment \cite{vezhnevets2017feudal} and guarantees balance between immediate and future revenue.

\minisection{Heterogeneous \& Variant Action Space}
Traditional RL models require a fixed action space \cite{mnih2013playing}.
If we model picking an order as an RL action, there is no guarantee of a fixed action space as the available orders keep changing. 
\citet{zhang2017taxi} proposed to learn a state-action value function to evaluate each valid order-vehicle match, then use a combinatorial optimization method such as Kuhn-Munkres (KM) algorithm \cite{munkres1957algorithms} to filter the matches.
However, such a method faces another important challenge that order dispatching and fleet management are different tasks, which results in heterogeneous action spaces.
To address this issue, we redefine action as the weight vector for ranking orders and fleet management, where the fleet controls are regarded as fake orders, and all the orders are ranked and matched with vehicles in each agent.
Thus, it bypasses the issue of heterogeneous and variant action space as well as high computational costs.

\minisection{Multi-Scale Ride-Hailing}
\citet{xu2018large} introduced a policy evaluation based RL method to learn the dynamics for each grid. 
As its result shows, orders and vehicles often centralize at different districts (e.g. uptown and downtown in Figure~\ref{fig:thermodynamics}). 
How to combine large hotspots in the city (inter-district) with small ones in districts (intra-district) is another challenge without much attention.
In order to take both inter-district and intra-district allocation into consideration, we adopt and extend attention mechanism in a hierarchical way (as will be further discussed in Figure~\ref{fig:coRide}). 
Compared with learning value function for each grid homogeneously \cite{xu2018large}, this attention-based structure can not only capture the impacts of neighbor agents, but also learn to distinguish the key grid and district in \emph{worker} (grid) and \emph{manager} (district) scales respectively.\par

Wrapping all modules together, we propose \emph{CoRide}, a hierarchical multi-agent reinforcement learning framework to resolve the aforementioned challenges. 
The main contributions are listed as follows:\par

\begin{itemize}[topsep = 3pt,leftmargin =5pt]
\item 	We propose a novel framework that learns to collaborate in hierarchical multi-agent setting for ride-hailing platform. 
\item	We conduct extensive experiments based on real-world data of multiple cities, as well as analytic synthetic data, which demonstrate that \emph{CoRide} provides superior performance in terms of ADI and ORR in the task of city-wide hybrid order dispatching and fleet management over strong baselines.
\item	To the best of our knowledge, \emph{CoRide} is the first work (i) to apply the hierarchical reinforcement learning on ride-hailing platform; (ii) to address the task of joint order dispatching and fleet management of online ride-hailing platforms; (iii) to introduce and study multi-scale ride-hailing task.
\end{itemize}

This structure conveys several benefits:
(i) In addition to balancing long-term and short-term reward, it also facilitates adaptation in a dynamic real-world situation by assigning different goals to \emph{worker}.  
(ii) Instead of considering all of the matches between available orders and vehicles globally, these tasks are distributed to each \emph{worker} and \emph{manager} agent and fulfilled in a parallel way.\par

\section{Related Work}
\label{related}
\minisection{Decision-making for Ride-hailing}
Order dispatching and fleet management are two major decision-making tasks for online ride-hailing platforms, which have acquired much attention from academia and industry during the recent few years.\par

To tackle these challenging transportation problems, automatically ruled-based approaches addressed order dispatching problem with either centralized or decentralized settings.
To improve global performance, \citet{zhang2017taxi} proposed a novel model based on centralized combinatorial optimization by concurrently matching multiple vehicle-order pairs within a short time window. 
However, this approach needs to compute all available vehicle-order matches and requires feature engineering, which would be infeasible and prevent it to be adopted in the large-scale taxi-order dispatching situation. 
With the decentralized setting, \citet{seow2010collaborative} addressed this problem with a collaborative multi-agent taxi dispatching system. 
However, this method requires rounds of direct communications between agents, so it is limited to a local area with a small number of vehicles. \par

Instead of rule-based approaches, which require additional handcrafted heuristics, the current trending direction is to incorporate reinforcement learning algorithms in complicated traffic management problems. 
\citet{xu2018large} proposed a learning and planning method based on reinforcement learning to optimize vehicle utilization and user experience in a global and more farsighted view. 
In \cite{oda2018distributed}, the authors leveraged the graph structure of the road network and expanded distributed DQN formulation to maximize entropy in the agents' learning policy with soft Q-learning, to improve performance of fleet management. 
\citet{wei2018look} introduced a reinforcement learning method, which takes the uncertainty of future requests into account and can make a look-ahead decision to help the operator improve the global level-of-service of a shared-vehicle system through fleet management. 
To capture the complicated stochastic demand-supply variations in high-dimensional space, \citet{lin2018efficient} proposed a contextual multi-agent actor-critic framework to achieve explicit coordination among a large number of agents adaptive to different contexts in fleet management system.\par

Different from all aforementioned methods, our approach is the first, to the best of our knowledge, to consider the joint modeling of order dispatching and fleet management and also the only current work introducing and studying the multi-scale ride-hailing task.\par

\minisection{Hierarchical Reinforcement Learning}
Hierarchical reinforcement learning (HRL) is a promising approach to extend traditional reinforcement learning (RL) methods to solve tasks with long-term dependency or multi-level interaction patterns \cite{dayan1993feudal,dietterich2000hierarchical}. 
Recent works have suggested that several interesting and standout results can be induced by training multi-level hierarchical policy in a multi-task setup \cite{frans2017meta,sigaud2018policy} or implementing hierarchical setting in sparse reward problems \cite{riedmiller2018learning,vezhnevets2017feudal}.\par

The options framework \cite{stolle2002learning,precup2000temporal,sutton1999between} formulates the problem with a two-level hierarchy, where the low-level - option - is a sub-policy with a termination condition. 
Since traditional options framework suffers from prior knowledge on designing options, jointly learning high-level policy with low-level policy has been proposed by \cite{bacon2017option}. 
However, this actor-critic HRL approach needs to either learning sub-policies for each time step or one sub-policy for the whole episode. Therefore, the performance of the whole module often prone to learning useful sub-policies. To guarantee gaining effective sub-policies, one alternative direction is to provide auxiliary rewards for low-level policy: hand-designed rewards based on prior domain knowledge \cite{kulkarni2016hierarchical,tessler2017deep} or mutual information \cite{florensa2017stochastic,daniel2012hierarchical,kong2017effective}. 
Given having access to one well-designed and suitable reward is often a luxury, \citet{vezhnevets2017feudal} proposed FeUdal Networks (FuN), where a generic reward is utilized for low-level policy learning, thus avoid designing hand-craft rewards. 
Several works extend and improve FuN with off-policy training \cite{nachum2018data}, form of hindsight \cite{levy2018learning} and representation learning \cite{nachum2018near}.\par

Our work is also developed from FuN \cite{vezhnevets2017feudal}, originally inspired by feudal RL \cite{dayan1993feudal}.
FuN employs only one pair of \emph{manager} and \emph{worker} and connects them with a parameterized goal and intrinsic reward.
Instead, we model \emph{CoRide} with multiple \emph{manager}s.
Unlike our method, in FuN the \emph{manager} and \emph{worker} modules are set to one-to-one, share the same observation, and operate at the different temporal but same spatial resolution.
In \emph{CoRide}, there are multiple \emph{worker}s learning to collaborate under one \emph{manager} while the \emph{manager}s are also coordinating. 
The \emph{manager} takes a joint observation of all \emph{worker}s, and each \emph{worker} produces action based on specific observation and sharing goal.
In other words, FuN \cite{vezhnevets2017feudal} is actually a special case of \emph{CoRide}, where a single \emph{manager} along with its only \emph{worker} is employed. 
Stepping on this one-to-many setting, the \emph{manager} can not only operate with long timescale credit but act at a lower spatial resolution.
Recently, \citet{ahilan2019feudal} introduced a novel architecture named FMH for cooperation in multi-agent RL. 
Different from this proposed method, \emph{CoRide} not only extends the scale of the multi-agent environment but also facilitates communication through multi-head attention mechanism, which computes influences of interactions and differentiates the impacts to each agent.
Yet, the majority of current HRL methods require careful task-specific design, making them difficult to apply in real-world scenarios \cite{nachum2018data}. 
To the best of our knowledge, \emph{CoRide} is the first work to apply hierarchical reinforcement learning on the ride-hailing problem.\par

\section{Problem Formulation}
\label{problem}
\begin{figure}[t]
\centering
\includegraphics[width=0.4\textwidth]{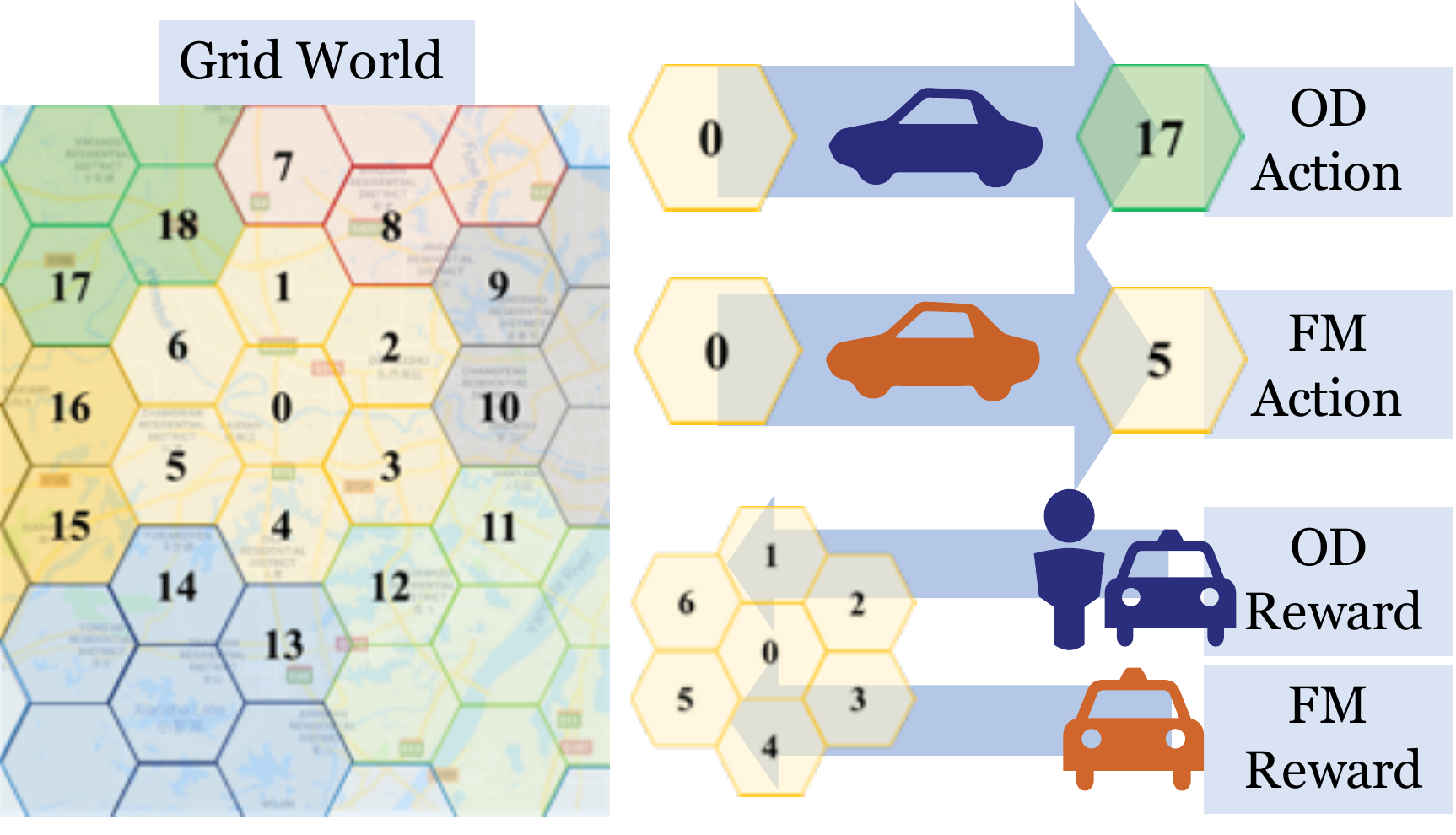}
\vspace{-3mm}
\caption {Illustration of the grid world and problem setting.}
\label{fig:grid}
\vspace{-3mm}
\end{figure}

We formulate the problem of controlling large-scale homogeneous vehicles in online ride-hailing platforms, which combines order dispatching system with fleet management system with the goal of maximizing the city-level ADI and ORR. 
In practice, vehicles are divided into two groups: order dispatching (OD) group and fleet management (FM) group. 
For OD group, we match these vehicles with available orders pair-wisely; whereas for FM group, we need to reposition them to the locations or dispatch orders to them (same as OD group). 
The illustration of the problem is shown in Figure~\ref{fig:grid}. 
We use the hexagonal-grid world to represent the map and take a grid as an agent. 
Considering that only orders within the pick-up distance can be dispatched to vehicle, we set distance between grids based on the pick-up distance. 
Given that, in our setting, vehicles in the same spatial-temporal node are homogeneous, i.e., the vehicles located at the same grid share the same setting. 
As such, we can model order dispatching as a large-scale parallel ranking task, where we rank orders and match them with homogeneous vehicles in each grid. 
The fleet control for fleet management, i.e. repositioning the vehicle to neighbor grids or staying at the current grid, is treated as fake orders (as will be further discussed in Section~\ref{experiment}) and conducted in the ranking procedure as same as order dispatching.\par

Since each agent can only reposition vehicles located in the managing grid, we propose to formulate the problem using \emph{Partially Observable Markov Decision Process (POMDP)} \cite{spaan2012partially} in a hierarchical multi-agent reinforcement learning setting for both order dispatching and fleet management. 
Thus, we decompose the original complicated tasks into many local ones and transform a high-dimensional problem into multiple low-dimensional problems.\par 

Formally, we model this task as a Markov game $\mathcal{G}$ for $N$ agents, which is defined by a tuple $\mathcal{G} = (N,\mathcal{S},\mathcal{A},\mathcal{P},\mathcal{R},\gamma)$, where $N,\mathcal{S}$, $\mathcal{A}$, $\mathcal{P}$, $\mathcal{R}$, $\gamma$ are the number of agents, set of states, set of actions, state transition probability, reward function, and a future reward discount factor, respectively. 
The definitions of major components are as follows.\par

\minisection{Agent}
We consider each region cell as an agent identified by $i \in  \mathcal{I}$, where $ \mathcal{I} = \{i \mid i=1,\dots,N\}$. In detail, a single grid represents a \emph{worker} agent, a district containing multiple grids represents a \emph{manager} agent. 
An example is presented in Figure~\ref{fig:grid}. 
Each individual grid serves as a \emph{worker} agent with 6 neighbor grids, as shown in the same color, composes a \emph{manager} agent. Note that although the number of vehicles and orders varies over time, the number of agents is fixed.

\minisection{State $s_t \in  \mathcal{S}$, Observation $o_t \in  \mathcal{O}$}
Although there are two different types of agents - \emph{manager} and \emph{worker}, their observations only differ in scale. 
Observation of each \emph{manager} is actually a joint observation of its \emph{worker}s. 
At each timestep $t$, agent $i$ draws private observations $o^i_t \in  \mathcal{O}$ correlated with the environment state $s_t \in  \mathcal{S}$. 
In our setting, the state input used in our method is expressed as $\mathcal{S} = \langle N_{v}, N_{o}, E, N_{f}, D_{o}\rangle$, where inner elements represent the number of vehicles, number of order, entropy, number of vehicles in FM group and distribution of order features (e.g. price, duration) in current grid respectively. 
Note that both dispatching and repositioning belong to resource allocation similar to the thermodynamic system (Figure~\ref{fig:thermodynamics}), and once order dispatching or fleet management occurs, dispatched or fleeted items slip out of the system. 
Namely, only idle vehicles and available orders can contribute to disorder and unevenness of the ride-hailing system. 
Therefore, we introduce and extend the concept of entropy here, defined as:
\begin{equation}
E = -k_B \cdot \sum_i \rho_i \log \rho_i := -k_B \cdot \rho_0 \log \rho_0
\end{equation}
where $k_B$ is a Boltzmann constant, and $\rho_i$ means probability for each state: $\rho_1$ for dispatched and fleeted, $\rho_0$ elsewhere. 
As aforementioned analysis, once order and vehicle combined as order-vehicle pairs, both order and vehicle are out of the ride-hailing platform.
Therefore, we choose to ignore items at state 1 ($\rho_1$) and compute $\rho_0$ as proportion of available order-vehicle pairs in all potential pairs:
\begin{equation}
\rho_0 = \frac{\text{min}(N_{v}, N_{o}) \times \text{min}(N_{v}, N_{o})}{N_{v} \times N_{o}} = \frac{N_{v} \times N_{v}}{N_{v} \times N_{o}} = \frac{N_{v}}{N_{o}}
\end{equation}
which is conditioned in $N_{v} < N_{o}$ situation and straightforward to transform to other situations.\par

\minisection{Action $a_t \in  \mathcal{A}$, State Transition $\mathcal{P}$}
In our hierarchical RL setting, \emph{manager}'s action is to generate abstract and intrinsic goal to its \emph{worker}s, and each \emph{worker} needs to provide a ranking list of relevant real orders (OD) and fleet control served as fake orders (FM). 
Thus, the action of \emph{worker} is defined as the weight vector for the ranking features. 
Changing an action of the \emph{worker} means to change the weight vector for the ranking features (as will be further discussed in Section~\ref{experiment}). 
Each timestep the whole multi-agent system produces a joint action for each \emph{manager} and \emph{worker} $a_t \in \mathcal{A}$, where $\mathcal{A} := \mathcal{A}_1 \times ... \times \mathcal{A}_N$, which induces a transition in the environment according to the state transition $\mathcal{P}(s_{t+1}|s_{t}, a_{t})$.\par 

\minisection{Reward $\mathcal{R}$}
Like previous hierarchical RL settings \cite{vezhnevets2017feudal}, only \emph{manager} interacts with the environment and receives feedback from it. 
This extrinsic reward function determines the direction of optimization and is proportional to both immediate profit and potential value; while the intrinsic reward is set to encourage the \emph{worker} to follow the instruction from the \emph{manager}.
The details will be further discussed in Eq.~(\ref{mreward}) and Eq.~(\ref{wreward}).\par 

More specifically, we give a simple example based on the above problem setting in Figure~\ref{fig:grid}.
At time $t = 0$, the \emph{worker} agent 0 ranks available real orders and potential fake orders for fleet control, and selects the Selected-2 (as will be further discussed in Eq.~(\ref{selected})) options: a real order from grid 0 to grid 17, a fake order from grid 0 to grid 5. 
After the driver finished, the \emph{manager} agent, whose sub-\emph{worker}s maintain the \emph{worker} agent 0, received corresponding reward.\par

\begin{figure}[b]
\centering
\vspace{-3mm}
\includegraphics[width=0.45\textwidth]{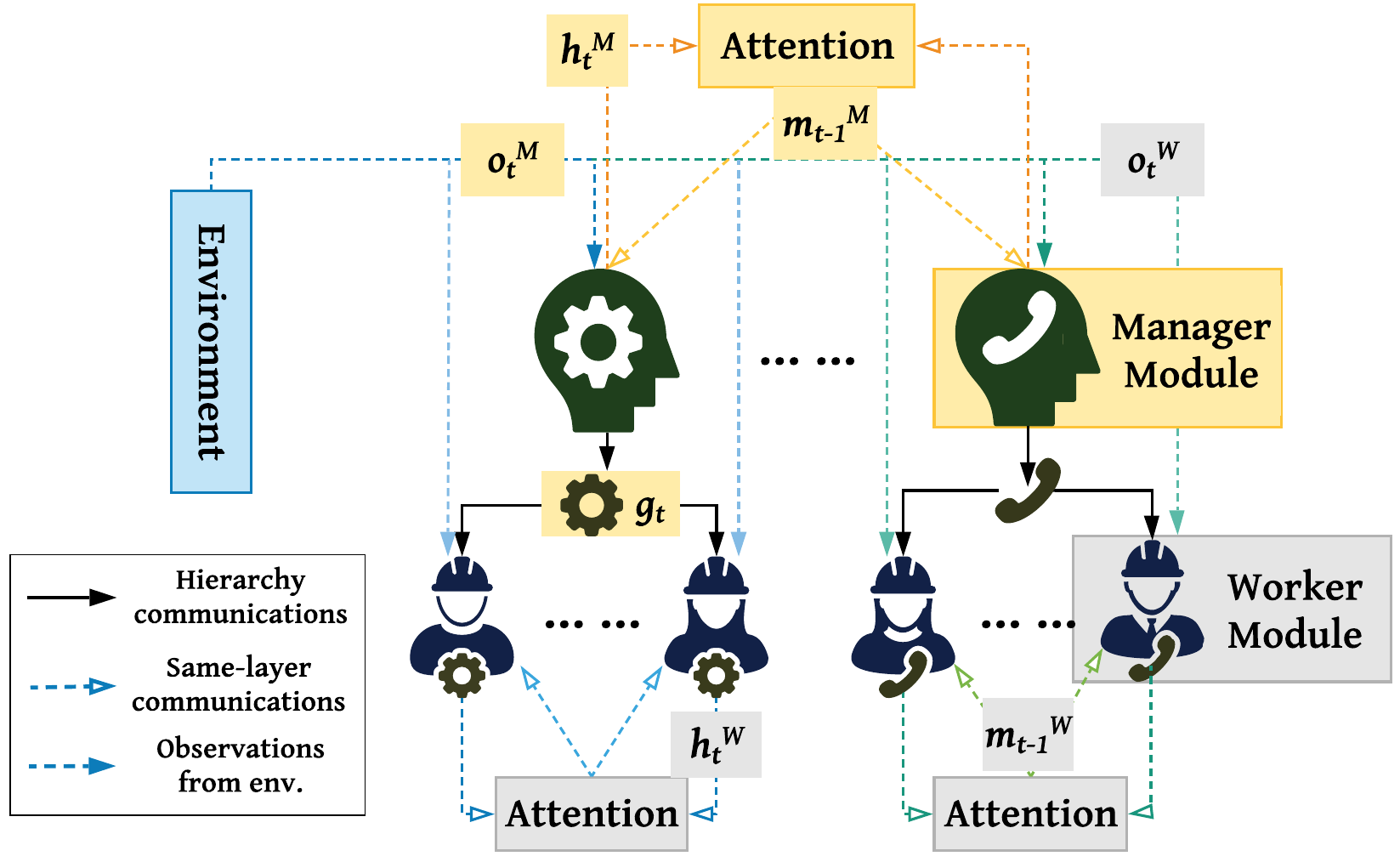}
\vspace{-3mm}
\caption {Overall architecture of CoRide.}
\label{fig:coRide}
\vspace{-3mm}
\end{figure}

\section{Methodologies}
\label{coride}
\subsection{Overall Architecture}
As shown in Figure~\ref{fig:coRide}, \emph{CoRide} employs two layers of agents, namely the layer of \emph{manager} agents and the layer of \emph{worker} agents.
Each agent is associated with a communication component for exchanging messages. 
As both agent and decision-making process conduct in a hierarchical way, multi-head attention mechanism served for communication is extended into multi-layer setting.\par

Compared with traditional one-to-one \emph{manager}-\emph{worker} control in hierarchical RL \cite{vezhnevets2017feudal}, we design one-to-many \emph{manager}-\emph{worker} pattern, extend the scale, and learn to collaborate on two layers of agents. 
The \emph{manager} internally computes a latent state representation $h_{t}^{M}$ as an input to the \emph{manager}-level attention network, and outputs a goal vector $g_t$. 
The \emph{worker} produces action and input for \emph{worker}-level attention conditioned on its private observation $o_t^W$, peer message $m_{t-1}^{W}$,  and the \emph{manager}'s goal $g_t$. 
The \emph{manager}-level and \emph{worker}-level attention networks share the same architecture introduced in Section~\ref{multi-attention}. 
The details and training procedure for \emph{manager} and \emph{worker} are given in following parts.\par

\begin{figure}[t]
\centering
\includegraphics[width=0.35\textwidth]{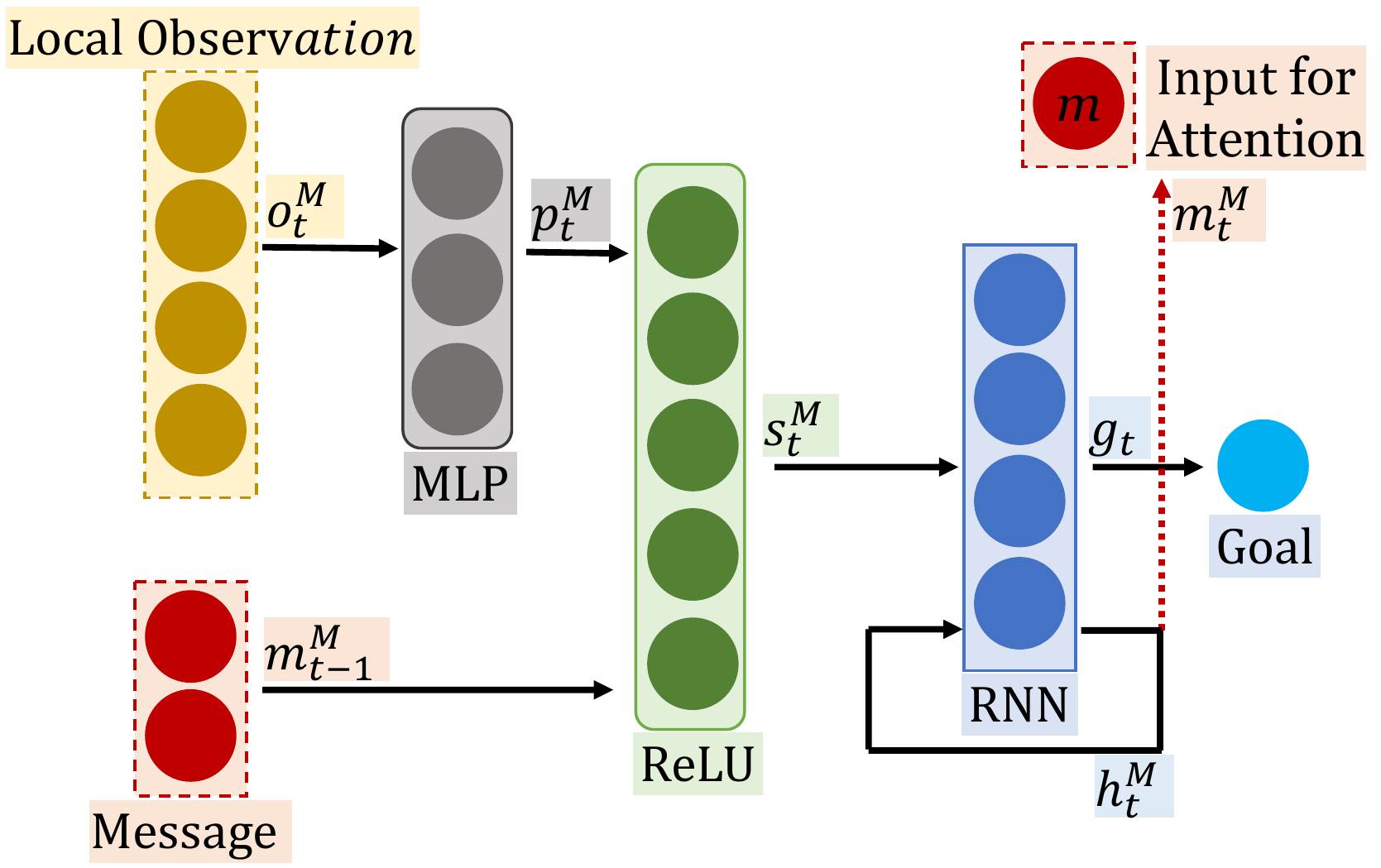}
\vspace{-3mm}
\caption {Manager Module.}
\label{fig:manager}
\vspace{-3mm}
\end{figure}

\subsection{Manager Module}
\label{manager}
The architecture of the \emph{manager} module is presented in Figure~\ref{fig:manager}. 
The \emph{manager} network is a two layer Preceptron (MLP) and a dilated RNN \cite{vezhnevets2017feudal}. 
Note that the structure of \emph{CoRide} and formula of the RNN enable \emph{manager} operate both at lower spatial resolution via taking joint observation of its \emph{worker}s and lower temporal resolution via dilated convolutional network \cite{yu2015multi}.\par

At timestep $t$, the agent receives an observation $o^M_t$ from the environment and feeds into the dilated RNN with peer messages $m^{M}_{t-1}$. 
Goal $g_t$ and input for \emph{manager}-level attention $h^M_{t}$ are generated as output of the RNN, governed by the following equations:
\begin{equation}
\label{goal}
h^{M}_{t}, \widehat{g}_t = \text{RNN}(s^M_t, h^{M}_{t-1}; \theta ^{rnn});  \quad
g_t = \widehat{g}_t/||\widehat{g}_t||
\end{equation}
where $\theta ^{rnn}$ is the parameters of the RNN network.
The environment responds with a new observation $o^M_{t+1}$ and a scalar reward $r_t$. 
The goal of the agent is to maximize the discounted return $\mathcal{R}_t = \sum_{k=0}^{\infty}\gamma^k r^M_{r+k+1}$ with $\gamma \in [0,1]$. Specifically, in the ride-hailing setting, we design our global reward taking both ADI and ORR into account, which can be formulated as:
\begin{equation}
\label{mreward}
r^M_t = r_{ADI} + r_{ORR}
\end{equation}
where $r_{ADI}$ denotes accumulated driver income, computed according to the price of each served order; while $r_{ORR}$ encourages ORR, and is calculated with several correlative factors as:
\begin{equation}
\label{ORR}
r_{ORR} = \sum_{grid} (E - \bar{E})^2 + \sum_{area} D_{KL}(\mathcal{P}^o_{t}\|\mathcal{P}^v_t)
\end{equation}
where $E, \bar{E}$ are the \emph{manager}'s entropy and global average entropy respectively.
Area, different from grid, often means a certain region which needs to be taken more care of. 
In our experiment, we select several grids whose entropy largely differs from the average as the area. 
$D_{KL}(\mathcal{P}^o_{t}\|\mathcal{P}^v_t)$ denotes Kullback-Leibler (KL) divergence which shows the margin between vehicle and order distributions of certain area at timestep $t$.
$\mathcal{P}^o_{t}$ and $\mathcal{P}^v_t$ are realized with Poisson distribution, a common distribution for vehicle routing \cite{ghiani2003real} and arriving \cite{lord2005poisson}.
In practice, this distribution parameters can be estimated from real trip data by the mean and std of orders and vehicles in each grid at each timestep.
Such a combined ORR reward design helps optimization both globally and locally.

\subsection{Worker Module}
\label{workermodel}
\begin{figure}[h]
\centering
\includegraphics[width=0.45\textwidth]{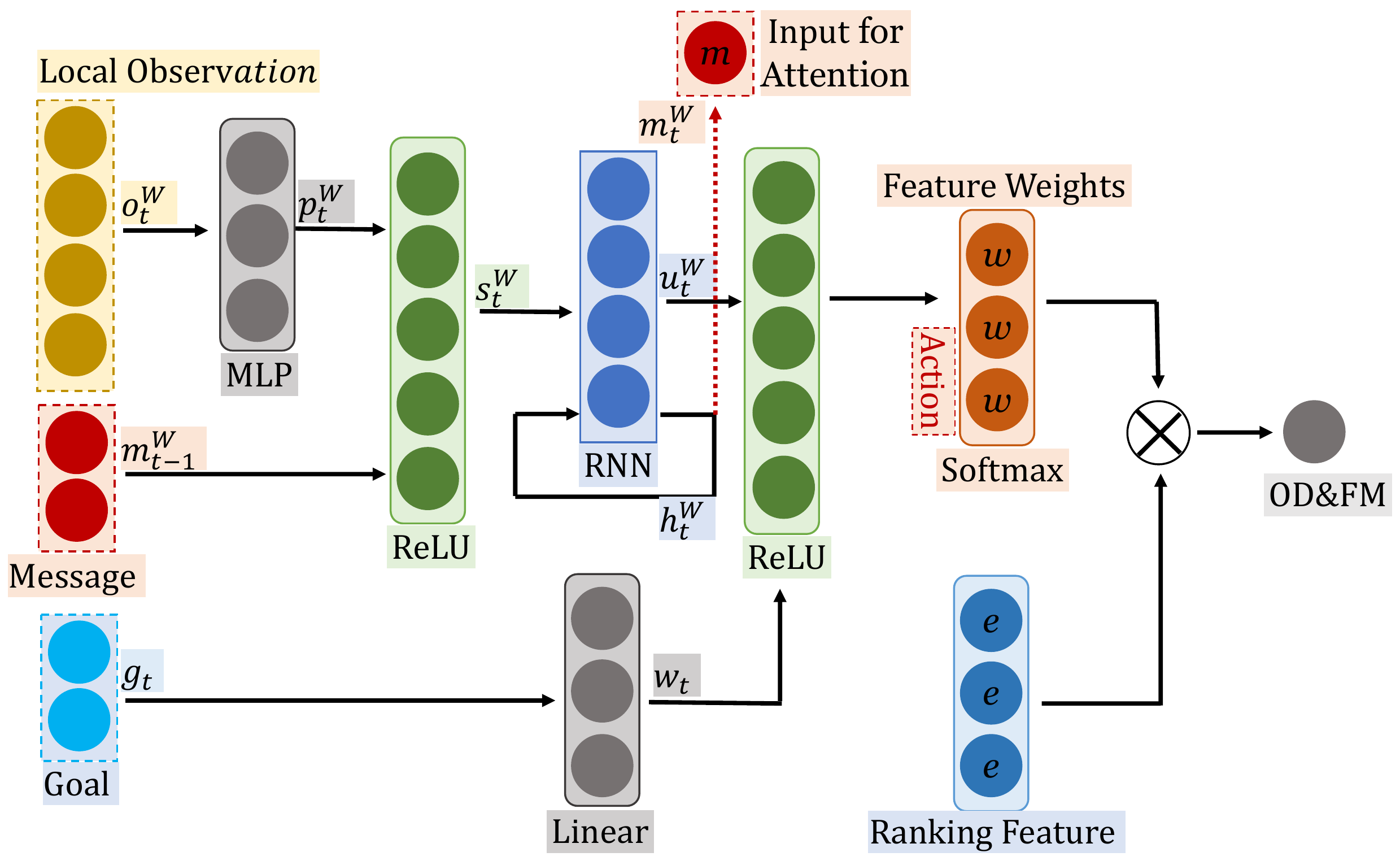}
\vspace{-3mm}
\caption {Worker Module.}
\label{fig:worker}
\vspace{-3mm}
\end{figure}

We adopt the goal embedding from Feudal Networks (FuN) \cite{vezhnevets2017feudal} in our \emph{worker} framework (see Figure~\ref{fig:worker}), where $w_t$ is generated as goal-embedding vector via linear projection. 
At each timestep $t$, the agent receives an observation $o_t^W$ from the environment and feeds into a regular RNN with peer message $m^W_{t-1}$. 
As Figure~\ref{fig:worker} shows, the output of RNN $u_t^W$ together with $w_t$ generates the primitive action - ranking weight vector $\omega_t$. 

Given that \emph{worker} needs to be encouraged to follow the goal generated by its \emph{manager}, we adopt the intrinsic reward proposed by \cite{vezhnevets2017feudal}, defined as:
\begin{equation}
\label{wreward}
r^W_t = \frac{1}{c} \sum_{i=1}^c d_{cos}(o^W_t-o^W_{t-i}, g_{t-i}),
\end{equation}
where $d_{cos}(\alpha, \beta) = \alpha^T\beta / (|\alpha|\cdot|\beta|)$ is the cosine similarity between two vectors. 
Notice that $g_t$ now represents an advantageous direction in the latent state space at a horizon $c$ \cite{vezhnevets2017feudal}. 
Such intrinsic reward design would provide directional shift for $worker$s to follow.\par

Different from traditional FuN \cite{vezhnevets2017feudal}, procedure of our \emph{worker} module produces action consists of two steps: 
(i) parameter generating, and 
(ii) action generating, inspired by \cite{zhao2017deep}. 
We utilize state-specific scoring function $f_{\theta^W}$ in parameter generating setup to map the current state $o^W_t$ to a list of weight vectors $\omega_t$ as:
\begin{equation}
\label{map}
f_{\theta^W}: o^W_t \rightarrow \omega_t, h^W_t
\end{equation}
which is calculated based on nerual network shown in Figure~\ref{fig:worker}.
In action generating setup, note that it is straightforward to extend linear relations with non-linear ones, we formulate that the scoring function parameter $\omega_t$ and the ranking feature $e_i$ for order $i$ as:
\begin{equation}
\label{rank}
score_{i} = \omega_t^T e_i
\end{equation} 
The detailed formulation of $e_i$ will be discussed in Section~\ref{simulator}.  
Then, we build and add real orders and potential fleet control - repositioning to neighbor grids and staying at the current grid - as fake orders into item space $\mathcal{I}$. 
After computing scores for all available options in $\mathcal{I}$, instead of directly ranking and selecting Top-$k$ items for order dispatching and fleet management, we adopt Boltzmann softmax selector to generate Selected-$k$ items:
\begin{equation}
\label{selected}
\text{Selected-}k = \frac{\text{exp}(score_i/\tau)}{\sum^M_{i=1}\text{exp}(score_i/\tau)} 
\end{equation}
where $k = \text{min}(N_{v}, N_{o})$, $\tau$ denotes temperature hyper-parameter to control the exploration rate, and $M$ is the number of scored order candidates. 
In practice, we set the initial temperature as 1.0, then gradually reduce the temperature until 0.01 to limit exploration. 
This approach not only equips the action selection procedure with controllable exploration but also diversify the policy's decision to avoid choosing groups of vehicles fleeted to the same destination.\par

\begin{algorithm}[!h]
\caption{\emph{CoRide} for joint multi-scale OD \& FM}
\label{framework}
\begin{algorithmic}[1]
\REQUIRE
current observations $o^M_t, o^W_t$; mutual communication messages $m_{t-1}^M, m_{t-1}^W$.
\vspace{1mm}
\FOR {each \emph{manager} in grid world}
\STATE Generate $g_t, h_t^M$ according to Eq.~(\ref{goal}).
\FOR {each \emph{worker} of the \emph{manager}}
	\STATE Generate $\omega_t, h_t^W$ according to Eq.~(\ref{map}).
	\STATE Add orders and fleet control items to item space $\mathcal{I}$.
	\STATE Rank items in $\mathcal{I}$ according to Eq.~(\ref{rank}).
	\STATE Generate Selected-$k$ items according to Eq.~(\ref{selected}).
\ENDFOR
\STATE $worker$-level attention mechanism generates $m_t^W$ according to Eq.~(\ref{message}).
\STATE \emph{manager} receives extrinsic reward $r^M_t$, and its \emph{worker}s receive intrinsic reward $r^W_t$ according to Eq.~(\ref{mreward}) and Eq.~(\ref{wreward}) respectively. 
\ENDFOR
\STATE $manager$-level attention mechanism generates $m_t^M$ according to Eq.~(\ref{message}).
\STATE Update parameters according to Algorithm~\ref{training}.
\end{algorithmic}
\end{algorithm}

\subsection{Multi-head Attention for Coordination}
\label{multi-attention}
Note that \emph{manager} and \emph{worker} share the same setting of multi-head attention mechanism, agent in this subsection can represent either of them. 
We utilize $h^i_{t-1}$ to denote the cooperating information for $i$-th agent generated from RNN at $t$-1, and extend self-attention mechanism to learn to evaluate each available interaction as: 
\begin{equation}
h^{ij}_{t-1} = (h^i_{t-1}W_T) \cdot (h^j_{t-1}W_S)^T
\end{equation}
where $h^i_{t-1}W_T, h^j_{t-1}W_S$ are embedding of messages from target agent and source agent respectively. 
We can model $h_{t-1}^{ij}$ as the value of communication between $i$-th agent and $j$-th agent. 
To retrieve a general attention value between source and target agents, we further normalize this value in neighborhood scope as:
\begin{equation}
\alpha_{t-1}^{ij} = \text{softmax}(h_{t-1}^{ij}) = \frac{\text{exp}(h_{t-1}^{ij} / \iota)}{\sum_{j \in \mathcal{N}_{i}} \text{exp}(h_{t-1}^{ij} / \iota)}
\end{equation}
where $\mathcal{N}_{i}$ is the neighborhood scope: the set of communication available for target agent, and $\iota$ denotes temperature factor. 
To jointly attend to the neighborhood from different representation subspaces at different grids, we leverage multi-head attention as in previous work \cite{vaswani2017attention,velivckovic2017graph,wei2019colight,zhang2019cityflow} to extend the observation as:
\begin{equation}
\label{message}
m_{t}^{i} = \sigma\Big(W_q \cdot \big(\cfrac{1}{H} \sum_{n=1}^{n=H} \sum_{j \in \mathcal{N}_{i}} \alpha_{t-1}^{ijn}(h_{t-1}^j W^n_C) \big)+b_q \Big)
\end{equation}
where $H$ is the number of attention heads, and {$W_T, W_S, W_C$} are multiple sets of trainable parameters. 
Thus, peer message $m_{t}$ is generated and will be feed into the corresponding module to produce the cooperative information $h_t$. 
We present the overall \emph{CoRide} for joint order dispatching and fleet management in Algorithm~\ref{framework}.\par

\begin{algorithm}[!b]
\caption{Parameters Training with DDPG}
\label{training}
\begin{algorithmic}[1]
\STATE Randomly initialize Critic network $Q(m_{t+1}, o_t, a_t | \theta^Q)$ and actor $\mu(m_{t-1}, o_t|\theta^\mu)$ with weights $\theta^Q$ and $\theta^\mu$.
\STATE Initialize target network $Q^\prime$ and $\mu^\prime$ with weights $\theta^{Q^\prime} \leftarrow \theta^{Q}$, $\theta^{\mu^\prime} \leftarrow \theta^{\mu}$.
\STATE Initialize replay buffer $R$.
\FOR {each training episode}
	\FOR {agent $i = 1$ to $M$}
		\STATE $m_0$ = initial message, $t = 1$.
		\WHILE {$t < T$ and $o_t \neq$ terminal}
			\STATE Select the action $a_t$ = $\mu_t(m_{t-1}, o_t| \theta^\mu)$ for active agent;
			\STATE Receive reward $r_t$ and new observation $o_{t+1}$;
			\STATE Generate message $m_t$ = Attention$(h^0_{t-1}, h^1_{t-1}, ..., h^K_{t-1})$, where $h^k_{t-1}$ is latent vector in RNN and $K$ denotes the number of neighboring agents;
		\ENDWHILE
	\STATE Store episode \{$m_0, o_1, a_1, r_1, m_1, o_2, a_2, r_2 ... $\} in $R$.
	\ENDFOR
	\STATE Sample a random minibatch of transitions $\mathcal{T}:$ $<m_{t-1},$ $o_{t},$ $a_t,$ $r_t,$ $o_{t+1}>$ from replay buffer $R$.
	\FOR {each transition $\mathcal{T}$}
		\STATE Set $y_t = r_t + \gamma Q^\prime (m_t, o_{t+1}, \mu(m_{t}, o_{t+1}|\theta^{\mu^\prime})| \theta^{Q^\prime})$;
		\STATE Update Critic by minimizing the loss:\\ 
		$L(\theta^{Q^\prime}) = (y_t - Q(m_{t-1}, o_t, a_t| \theta^Q))^2$;
		\STATE Update Actor policy by maximizing the Critic:\\ 
		$J(\theta^{\mu})= Q(m_{t-1}, o_t, a_t | \theta^Q)|_{a=\mu(m_{t-1}, o_t | \theta^\mu)}$;
		\STATE Update communication component.
	\ENDFOR
\ENDFOR
\end{algorithmic}
\end{algorithm}

\subsection{Training}
As described in Algorithm~\ref{framework}, \emph{manager}s generate specific goals based on their observations and peer messages (line 2). 
The \emph{worker}s under the \emph{manager} generate the weight vector according to private observation and sharing goal (line 4). 
We then build a general item space $\mathcal{I}$ for order dispatching and fleet management (line 5), and rank items in $\mathcal{I}$ (line 6). 
Considering that our action is conditional to the  minimum of the number of vehicles and orders, we generate Selected-$k$ items as the final action (line 7).\par

We extend learning approach from FuN \cite{vezhnevets2017feudal} and HIRO  \cite{nachum2018data} to train \emph{manager} and \emph{worker} module in the similar way. 
In \emph{CoRide}, we utilize DDPG algorithm \cite{lillicrap2015continuous} to train the parameters for both \emph{manager} and \emph{worker} module for following reasons. 
Classically, the critic is designed to leverage an approximator, to learn an action-value function $Q(o_t, a_t)$, and to direct the actor updating its parameters. 
The optimal action-value function $Q^*(o_t, a_t)$ should follow the Bellman equation \cite{bellman2013dynamic} as:
\begin{equation}
\label{traditional}
Q^* (o_t, a_t) = \mathbb{E}_{o_{t+1}}[r_t+\gamma \max_{a_{t+1}} Q^*(o_{t+1}, a_{t+1})|o_t, a_t]
\end{equation}
which requires $|\mathcal{A}|$ evaluations to select the optimal action. 
This prevents Eq.~(\ref{traditional}) to be adopted in real-world scenario, e.g. ride-hailing setting, with enormous state and action spaces. 
However, the actor architecture proposed in Section~\ref{workermodel} generates a deterministic action for critic. 
Furthermore, \citet{lillicrap2015continuous} proposed a flexible and practical method to use an approximator function to estimate the action-value function, i.e. $Q(o, a) \thickapprox Q(o, a; \theta^\mu)$. 
In practice, we refer to leverage DQN: a neural network function approximator can be trained by minimizing a sequence of loss functions $L(\theta^\mu)$ as:
\begin{equation}
L(\theta^\mu) = \mathbb{E}_{s_t, a_t, r_t, o_{t+1}}[(y_t - Q(o_t, a_t; \theta^\mu))^2]
\end{equation}
where $y_t = \mathbb{E}_{o_{t+1}}[r_t + \gamma Q^\prime(o_{t+1}, a_{t+1}; \theta^{\mu^\prime})|o_{t}, a_{t}]$ is the target for the current iteration.
According to the aforementioned analysis, the general training algorithm for the \emph{manager} and \emph{worker} module is presented in Algorithm~\ref{training}.

\begin{table*}[t]
\centering
\caption{Performance comparison of competing methods in terms of ADI and ORR with respect to the performance of RAN. 
For a fair comparison, the random seeds that control the dynamics of the environment are set to be the same across all methods.}
\vspace{-3mm}
\begin{tabular}{|c|cc|cc|cc|}
\hline
City & \multicolumn{2}{c|}{City A} & \multicolumn{2}{c|}{City B} & \multicolumn{2}{c|}{City C}\\
\hline
Metrics & Normalized ADI & Normalized ORR & Normalized ADI & Normalized ORR & Normalized ADI & Normalized ORR\\
\hline
\hline
DQN & +5.71\% $\pm$ 0.02\% & +2.67\% $\pm$ 0.01\% & +6.30\% $\pm$ 0.01\% & +3.01\% $\pm$ 0.02\% & +6.11\% $\pm$ 0.02\% & +3.04\% $\pm$ 0.01\% \\
MDP & +7.11\% $\pm$ 0.05\% & +2.71\% $\pm$ 0.03\% & +7.89\% $\pm$ 0.05\% & +3.13\% $\pm$ 0.04\% & +7.53\% $\pm$ 0.03\% & +3.19\% $\pm$ 0.03\% \\
DDQN & +6.68\%  $\pm$ 0.04\% & +3.19\% $\pm$ 0.04\% & +7.75\% $\pm$ 0.06\% & +4.06\% $\pm$ 0.05\% & +7.62\% $\pm$ 0.04\% & +4.58\% $\pm$ 0.05\% \\
MFOD & +6.62\% $\pm$ 0.03\% & +3.71\% $\pm$ 0.02\% & +7.91\% $\pm$ 0.04\% & +4.01\% $\pm$ 0.02\% & +7.32\% $\pm$ 0.02\% & +4.60\% $\pm$ 0.01\% \\
\hline
\hline
CoRide- & +9.27\% $\pm$ 0.04\% & +4.23\% $\pm$ 0.03\% & +8.73\% $\pm$ 0.03\% & +4.35\% $\pm$ 0.02\% & +9.06\% $\pm$ 0.03\% & +4.23\% $\pm$ 0.04\%\\
CoRide & +\textbf{9.80}\% $\pm$ 0.04\% & +4.81\% $\pm$ 0.05\% & +\textbf{8.94}\% $\pm$ 0.06\% & +4.89\% $\pm$ 0.04\% & +\textbf{9.23}\% $\pm$ 0.05\% & +5.19\% $\pm$ 0.04\% \\
\hline
\end{tabular}
\label{restable}
\end{table*}

\section{Simulator}
\label{simulator}
The trial-and-error nature of reinforcement learning requires a dynamic simulation environment for training and evaluation. 
Thus, we adopt and extend the grid-based simulator designed by \citet{lin2018efficient} to joint order dispatching and fleet management.\par

\subsection{Data Description}
\label{data}
The real world data provided by Didi Chuxing$^\dagger$\blfootnote{$^\dagger$Similar dataset supported by Didi Chuxing can be found via GAIA open dataset (https://outreach.didichuxing.com/research/opendata/en/).}
includes order information and trajectories of vehicles in the central area of three large cities with millions of orders in four consecutive weeks. 
Data of each day contains million-level orders and tens of thousands vehicles in each city. 
The order information includes order price, origin, destination, and duration. 
The trajectories contain the positions (latitude and longitude) and status (on-line, off-line, on-service) of all vehicles every few seconds. 
As the radius of grid is approximate 1.3 kilometers, the central area of the city is covered by a hexagonal grids world consisting of 182, 126, 112 grids in three cities respectively. 
In order to adapt to the grid-based simulator, we utilize unique gridID to represent position information.\par

\subsection{Simulator Design}
In the grid-based simulator, the city is covered by a hexagonal grid-world as illustrated in Figure~\ref{fig:grid}. 
At each timestep $t$, the simulator provides an observation $o_t$ with a set of idle vehicles and a set of available orders including real orders and aforementioned fake orders for fleet control. 
All such fake orders share the same attributes as real orders, except that some of attributes are set stationary like price.
All these real orders are generated by bootstrapping from real-world dataset introduced above. 
More specifically, suppose the current timestep of simulator is $t$, we randomly sample real orders occuring in the same period, i.e. happening between $t_\Delta \times t$ to $t_\Delta \times (t+1)$, where $t_\Delta$ denotes timestep interval.
In practice, we set \emph{sampling rate} 100\%.
Like orders, vehicles are set online and offline alternatively according to a distribution learned from real-world dataset via maximum likelihood estimation. 
Each order feature, i.e. ranking feature $e_i$ in Eq.~(\ref{rank}), includes the origin gridID, the destination gridID, price, duration and the type of order indicating real or fake order; while each vehicle takes its gridID as a feature, and vehicles located at the same grid are regarded as homogeneous.
Moreover, as the travel distance between neighboring grids is approximately 1.3 kilometers and timestep interval $t_{\Delta}$ is 10 minutes, we assume that vehicles will not automatically move to other grids before taking another order.  
The ride-hailing platform then provides an optimal list of vehicle-order pairs according to current policy.
After receiving the list, the simulator will return a new observation $o_{t+1}$ and a list of order fees. 
Stepping on this feedback, rewards $r_t$ for each agent will be calculated and the corresponding record $(o_t, a_t, r_t, o_{t+1})$ will be stored into a replay buffer.
The whole network parameters will be updated using a batch of samples from replay buffer.\par

The effectiveness of the grid-based simulator is evaluated based on the calibration against the real data in term of the most important performance measurement: accumulated driver income (ADI) \cite{lin2018efficient}. 
The coefficient of determination $r^{2}$ between simulated ADI and real ADI is 0.9331 and the Pearson correlation is 0.9853 with $p$-value $p < 0.00001$.\par

\section{Experiment}
\label{experiment}
In this section, we conduct extensive experiments to evaluate the effectiveness of our proposed method in joint order dispatching and fleet management environment. 
Given that there are no published methods fitting our task. 
Thus, we first compare our proposed method with other models either widely used in the industry or published as academic papers based on a single order dispatching environment. 
Then, we further evaluate our proposed method in
joint setting and compare with its performance in single setting.\par

\subsection{Compared Methods}
\label{methods}
As discussed in \cite{lin2018efficient}, learning-based methods, currently regarded as state-of-the-art methods, usually outperform rule-based methods. 
Thus, we employ 6 learning-based methods and random method as the benchmark for comparison in our experiments.
\begin{itemize}[leftmargin=8pt]
\item \textbf{RAN}: A random dispatching algorithm considering no additional information. It only assigns idle vehicles with available orders randomly at each timestep.

\item \textbf{DQN}: 
\citet{li2019efficient} conducted action-value function approximation based on $Q$-network. 
The $Q$-network is parameterized by a MLP with four hidden layers and we adopt the ReLU activation between hidden layers and to transform the final linear output of $Q$-network. 
 
\item \textbf{MDP}: \citet{xu2018large} 
implemented dispatching through a learning and planning approach: each vehicle-order pair is valued in consideration of both immediate rewards and future gains in the learning step, and dispatch is solved using a combinatorial optimizing algorithm in planning step.
 
\item \textbf{DDQN}: \citet{wang2018deep} introduced a double-DQN with spatial-temporal action search.
The network architecture is similar to the one described in DQN except that a selected action space is utilized and network parameters are updated via double-DQN.

\item  \textbf{MFOD}: \citet{li2019efficient} modeled the order dispatching problem with MFRL \cite{pmlr-v80-yang18d} and simplified the local interactions by taking an average action among neighborhoods.


\item \textbf{CoRide}: Our proposed model as detailed in Section~\ref{coride}.

\item \textbf{CoRide-}: In order to further evaluate performance for hierarchical setting and agent communication, we set \emph{CoRide} without multi-head attention mechanism as one of the baselines.

\end{itemize}

\subsection{Result Analysis}
For all learning methods, following \cite{li2019efficient}, we run 20 episodes for training, store the trained model periodically, and conduct the evaluation on the stored model with 5 random seeds. We compare the performance of different models regarding two criteria, including ADI, computed as the total income in a day, and ORR, calculated by the number of orders taken divided by the number of orders generated.

\minisection{Experimental Results and Analysis}
As shown in Table~\ref{restable}, the performance surpasses the state-of-the-art models like DDQN and industry deployed model like MDP. 
DDQN along with DQN mainly limits in lack of interaction and cooperation in the multi-agent environment. 
MDP mainly focuses on order price but ignores other features of order like duration, which makes against finding a balance between getting higher income per order and taking more orders. 
Instead, our proposed method achieves higher growths in term of ADI not only by considering every feature of each order concurrently but through learning to collaborate hierarchically. 
MFOD trys to capture dynamic demand-supply variations by propagating many local interactions between vehicles and the environment among mean field. 
Note that the number and information of available grid are relatively stationary while the number and feature of active vehicles are more dynamic.
Thus, \emph{CoRide}, which takes grid as agent, is more likely and easier to learn to cooperate from interaction between agents.\par
\begin{figure}[h]
\centering
\vspace{-3mm}
\includegraphics[width=0.5\textwidth]{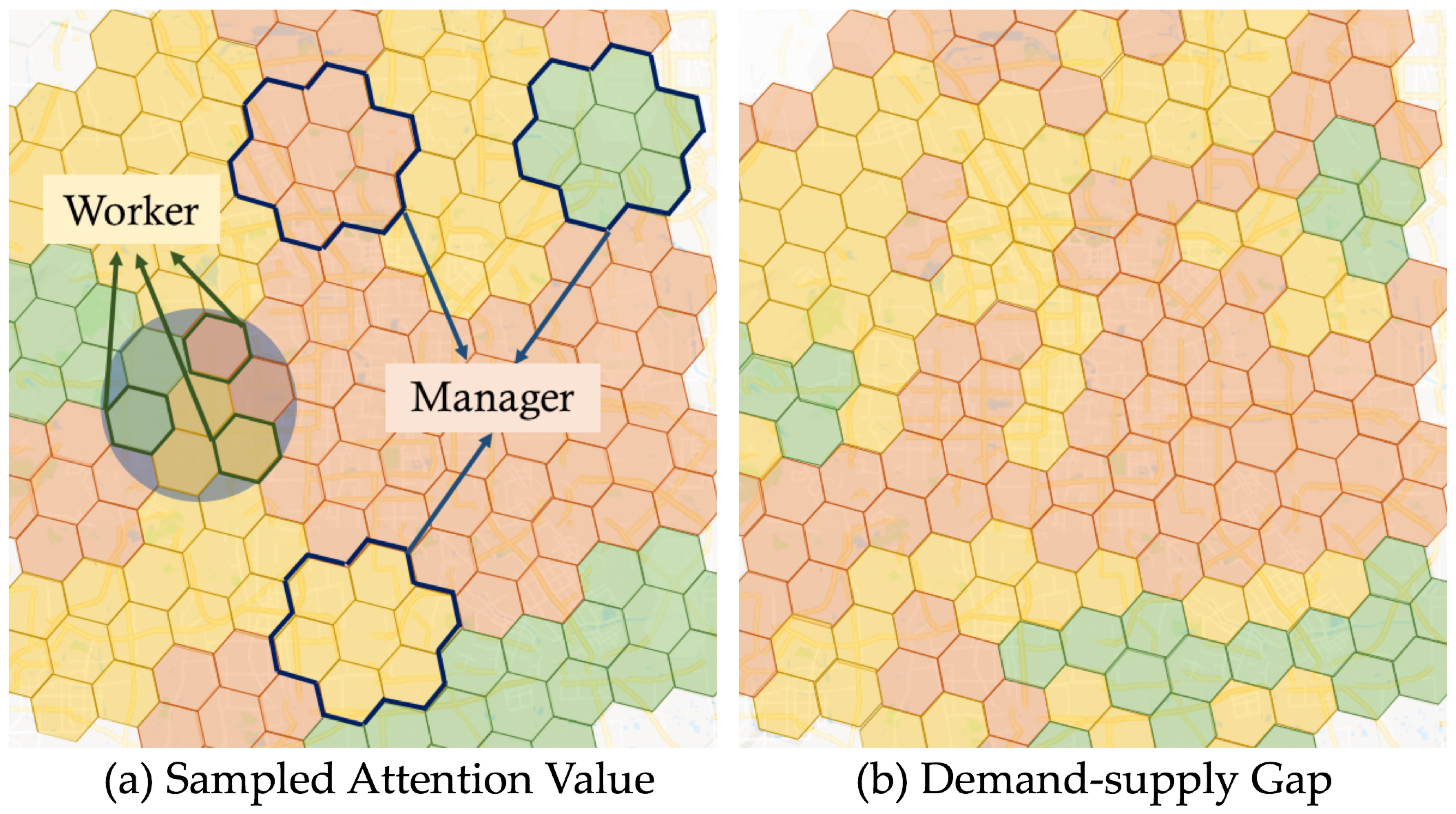}
\caption {Sampled attention value and demand-supply gap in the city center during peak hours. Grids with more orders or higher attention value are shown in red (in green if opposite) and the gap is proportional to the shade of colors.}
\vspace{-3mm}
\label{fig:attention}
\end{figure}

Apart from cooperation, multi-head attention network also enables \emph{CoRide} to capture demand-supply dynamics from both district (\emph{manager}) and grid (\emph{worker}) scale (as will be further discussed in Figure~\ref{fig:attention}). 
Such a novel combined scale setting facilitates  \emph{CoRide} both effectively and efficiently.\par

\minisection{Visualization Analysis} 
Except for quantitive results, we also analyze whether the learned multi-head attention network can capture the demand-supply relation (see Figure~\ref{fig:attention}(b)) through visualization. 
As shown in Figure~\ref{fig:coRide}, our communication mechanism conducts in a hierarchical way: attention among the \emph{manager}s communicates and learns to collaborate abstractly and globally while peers in \emph{worker}-layer operate and determine key grid locally.\par

The values of several example \emph{manager}s and a group of \emph{worker}s belonging to the same \emph{manager} are visualized in Figure~\ref{fig:attention}(a).
By taking a closer look at Figure~\ref{fig:attention}, we can observe that the area with high demand-supply indeed centralized at certain places, which has been well captured in \emph{manager}-scale.
Such district-level attention value allows precious vehicles to be dispatched efficiently in a global view.
Apart from \emph{manager}-scale one, multi-head attention network also provides \emph{worker}-scale attention value, which focuses on local allocation.
Stepping on this multi-scale dispatching system design, \emph{CoRide} could operate as a microscope, where coarse and fine focuses work together to obtain precise action.

\begin{table*}[t]
\centering
\caption{
Performance comparison of competing methods in terms of AST and TNF with three different \emph{discounted rate}s (DR). The numbers in Trajectory denote gridID in Figure~\ref{fig:case} and its color denotes the district it located in. O and W mean the vehicle is On-service and Waiting at the current grid. Also, we use underlined number to present fleet management.}

\vspace{-3mm}
\begin{tabular}{|c|ccc|ccc|ccc|}
\hline
DR & \multicolumn{3}{c|}{20\%} & \multicolumn{3}{c|}{30\%} & \multicolumn{3}{c|}{40\%}\\
\hline
Metrics & Trajectory & AST & TNF & Trajectory & AST & TNF & Trajectory & AST & TNF\\
\hline
\hline
RES & {\color{yellow} 13},{\color{green} 9},{\color{green} W},{\color{green} 14},{\color{green} W},{\color{yellow} 13},{\color{yellow} 8},{\color{red} 2},{\color{yellow} 7},{\color{yellow} 11} & 8 & 8 & {\color{yellow} 13},{\color{yellow} W},{\color{green} 14},{\color{green} W},{\color{green} W},{\color{yellow} 13},{\color{yellow} 8},{\color{yellow} W},{\color{yellow} 7},{\color{yellow} 11} & 6 & 6 & {\color{yellow} 13},{\color{yellow} W},{\color{green} 14},{\color{green} W},{\color{green} W},{\color{green} W},{\color{green} W},{\color{green} 19},{\color{green} O},{\color{green} 9} & 5 & 4 \\
REV & {\color{yellow} O},{\color{green} O},{\color{green} 15},{\color{green} W},{\color{green} 20},{\color{green} O},{\color{green} O},{\color{yellow} O},{\color{yellow} O},{\color{yellow} 11} & 9 & 3  & {\color{yellow} O},{\color{green} O},{\color{green} 15},{\color{green} W},{\color{green} W},{\color{green} W},{\color{green} O},{\color{yellow} O},{\color{yellow} O},{\color{yellow} 11} & 7 & 2 & {\color{yellow} O},{\color{green} O},{\color{green} 15},{\color{green} W},{\color{green} W},{\color{green} W},{\color{green} 20},{\color{green} W},{\color{green} O},{\color{green} 14} & 6 & 3 \\
\hline
\hline
CoRide & {\color{yellow} 13},{\color{green} 9},{\color{green} W},{\color{red} O},{\color{red} 0},{\color{red} 4},{\color{red} O},{\color{red} 2},{\color{red} O},{\color{red} 5} & \textbf{9} & \textbf{6} & {\color{yellow} 13},{\color{yellow} W},{\color{yellow} O},{\color{yellow} 11},{\color{yellow} W},{\color{yellow} W},{\color{yellow} O},{\color{red} O},{\color{red} 0},{\color{red} 5} & 7 & 4 & {\color{yellow} 13},{\color{yellow} W},{\color{yellow} W},{\color{yellow} O},{\color{yellow} 17},{\color{yellow} W},{\color{yellow} W},{\color{yellow} O},{\color{yellow} 0},{\color{red} 3} & 6 & 3 \\
CoRide+ & {\color{yellow} 13},{\color{green} 9},{\color{green} W},{\color{red} O},{\color{red} 0},{\color{red} 4},{\color{red} O},{\color{red} 2},{\color{red} O},{\color{red} 5} & \textbf{9} & \textbf{6} & {\color{yellow} \underline{8}},{\color{red} \underline{3}},{\color{red} 0},{\color{red} 2},{\color{red} O},{\color{red} 4},{\color{red} O},{\color{red} 2},{\color{red} O},{\color{red} 5} & \textbf{8} & \textbf{5} & {\color{yellow} \underline{8}},{\color{red} \underline{3}},{\color{red} 0},{\color{red} 2},{\color{red} O},{\color{red} 4},{\color{red} O},{\color{red} 2},{\color{red} O},{\color{red} 5} & \textbf{8} & \textbf{5}\\
\hline
\end{tabular}
\label{casetable}
\end{table*}

\minisection{Ablation Study}
In this subsection, we evaluate the effectiveness of components of \emph{CoRide}.
Notice that \emph{manager} and \emph{worker} modules serve as key components and are integrated through the hierarchical multi-agent architecture, as Figure~\ref{fig:coRide} shows. 
Thus, we choose to investigate the performance of multi-head attention network here and set \emph{CoRide-} as a variation of proposed method.
As shown in the last two rows in Table~\ref{restable}, \emph{CoRide-} achieves significant advantages over the aforementioned baselines, especially in City A.
Similar results occur with \emph{CoRide}.
This phenomenon can be explained from the fact that City A is the largest one according to Section~\ref{data}, which requires frequent and large numbers of transportations among regions.
Multi-scale guidance via multi-head attention network and  hierarchical multi-agent architecture is therefore potentially helpful, especially at a large-scale case.

\begin{SCfigure}
\centering
\includegraphics[width=0.65\columnwidth]{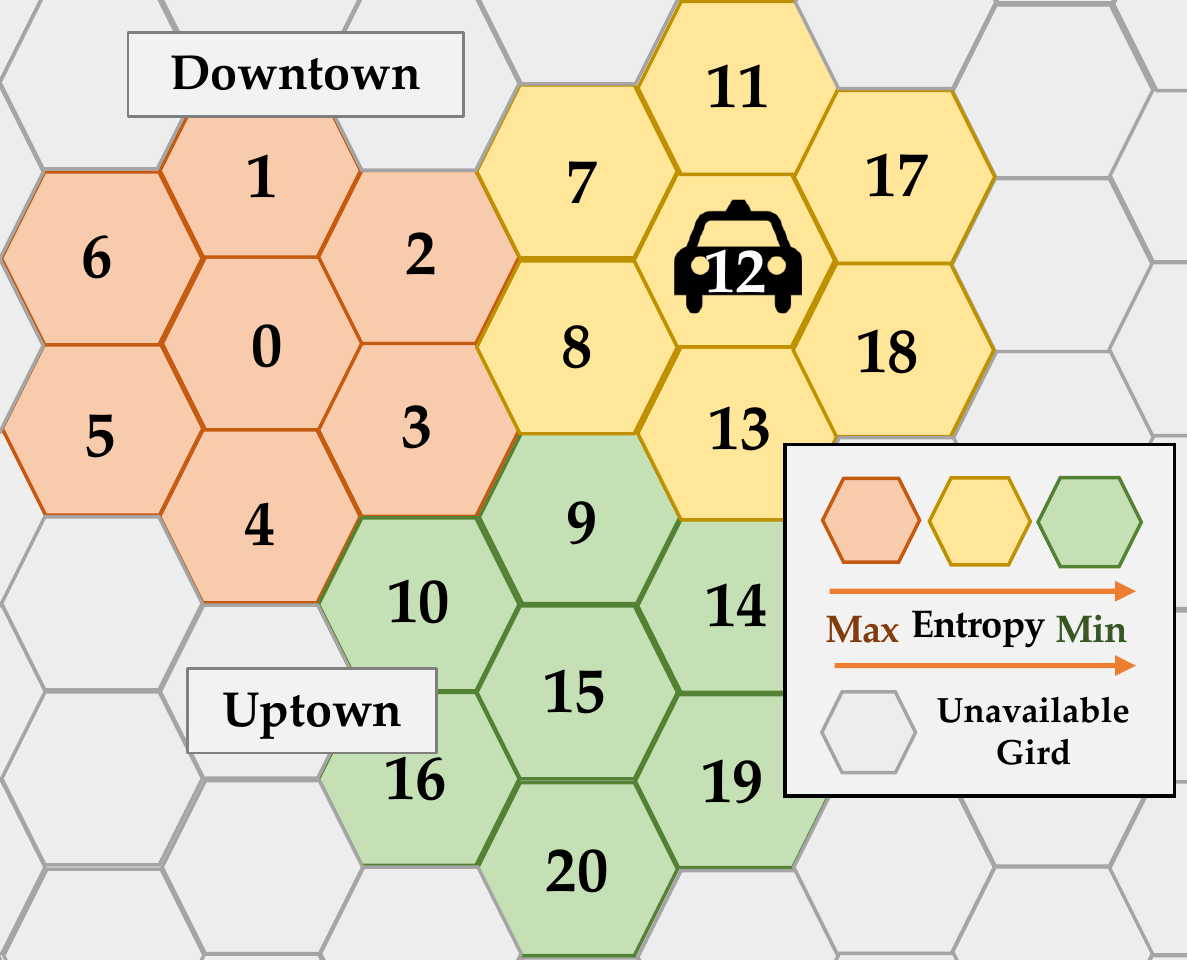}
\caption {Illustration of the grid world in case study, where color of the grids denote their entropy.
 }
\vspace{-5mm}
\label{fig:case}
\end{SCfigure}

\minisection{Case Study}
\label{case}
The above experimental results show that the success rate of our model is significantly better than others in single dispatching task. 
In order to evaluate the performance of \emph{CoRide} in joint dispatching and repositioning. 
Also, to further differ the formulations of the models, we constructed a synthetic dataset containing 3 districts with 21 grids, as showed in Figure~\ref{fig:case}.
All these synthetic datasets are obtained via sampling real-world dataset supported by Didi Chuxing. 
More concretely, order distributions of all grids are sampled from the average distribution in real world dataset.
Namely, order distributions in each grid are homogeneous.
In order to differ downtown areas from uptown areas, we introduce \emph{sampling rate} here.
The \emph{sampling rate} for each grid denotes popular rate in the real world.
We set downtown (red grids in Figure~\ref{fig:case}) with stationary \emph{sampling rate} 100\%.
The other regions are sampled with $\emph{sampling rate} := 100\% - \emph{discounted rate}$ for yellow district and $100\% - 2 \times \emph{discounted rate}$ for green district.
Specifically we set \emph{discounted rate} as $20\%$, $30\%$ and $40\%$ respectively, and further verify our proposed model by comparing against following 3 methods:\par

\begin{itemize}[leftmargin=8pt]
\item \textbf{RES}:  This response-based method aims to
achieve higher ORR, which corresponds to Total Number of Finished orders (TNF) in this section.
Orders with short duration will gain high priority to get dispatched first.
Once there are multiple orders with the same estimated trip time, then orders with higher prices will be served first.\par

\item \textbf{REV}:  The revenue-based algorithm focuses on a higher ADI, which corresponds to Accumulated on-Service Time (AST) in this section, by assigning vehicles to high price orders.
Following the similar principle as described above, the price and duration of orders will be considered as primary and secondary factors respectively.\par

\item \textbf{CoRide+}: To distinguish \emph{CoRide} running in different environment: single order dispatching, and joint order dispatching and fleet management, we sign the former one \emph{CoRide} and latter one \emph{CoRide+}.
\end{itemize}

In order to analyze these performances in a more straightforward way, we mainly employ rule-based methods here.
Also, we introduce AST calculated as the accumulated on-service time and TNF computed as the total number of finished orders as the new metrics corresponding to ADI and ORR respectively.
In order to further analyze these performances in a long-term way, we select one vehicle starting at grid 12, trace 10 timesteps and record its trajectory (as Figure~\ref{fig:case} shows), then conclude these results in Table~\ref{casetable}.
Although we only record the first 10 timesteps, we can observe that our proposed methods, both \emph{CoRide+} and \emph{CoRide}, are guiding the vehicle to regions with larger entropy.
This is benefit from architecture where the state of both \emph{manager} and \emph{worker}, and ranking feature $e_i$ take the grid information into consideration.
In contrast, other methods greedily optimize either AST (ADI) or TNF (ORR) and ignore these information.
After taking a close look at Table~\ref{casetable}, we can find that \emph{CoRide} and \emph{CoRide+} share the same trajectory on \emph{discounted rate} 20\% and differ greatly when \emph{discounted rate} moves to 30\% and 40 \%.
This can be explained by regarding \emph{CoRide+} as a combined design between our proposed model \emph{CoRide} and joint order dispatching and fleet management setting.
Namely, \emph{CoRide} is actually a special case of \emph{CoRide+}, where fleet management is unable. 
Equipped with fleet management, \emph{CoRide+} allows the vehicle move to and serve order in the hotspots more directly than \emph{CoRide}.
Also, when \emph{discounted rate} varies from 20\% to 40\%, fleet management enables \emph{CoRide+} with better adaptation and achieve stable performance, even can ignore the dynamics of order distributions in some cases. \par

According to aforementioned analysis, we can conclude that (i) \emph{CoRide+} achieves not only the state-of-the-art but a more stable result benefiting from joint order dispatching and fleet management setting; (ii) both \emph{CoRide} and \emph{CoRide+} can direct the vehicle to grids with larger entropy via taking grid information into consideration.

\section{Conclusion and future work}
\label{conclusion}
In this paper, we proposed \emph{CoRide}, a hierarchical multi-agent reinforcement learning solution to combine order dispatching and fleet management for multi-scale ride-hailing platforms. The results on multi-city real-world data as well as analytic synthetic data show that our proposed algorithm achieves (i) a higher ADI and ORR than aforementioned methods, (ii) a multi-scale decision-making process, (iii) a hierarchical multi-agent architecture in the ride-hailing task and (iv) a more stable method at different cases.
Note that \emph{CoRide} could achieve fully decentralized execution and incorporate closely with other geographical information based model like estimating time of arrival (ETA) \cite{wang2018learning} theoretically.
Thus, it's interesting to conduct further evaluation and investigation.
Also, we notice that applying hierarchical reinforcement learning in real-world scenarios is very challenge and our work is just a start. 
There is much work for future research to improve both stability and performance of hierarchical reinforcement learning methods on real-world tasks.

\minisection{Acknowledgments} 
The corresponding author Weinan Zhang thanks the support of National Natural Science Foundation of China (Grant No. 61702327, 61772333, 61632017). 
We would also like to thank our colleague in DiDi for constant support and encouragement.

\bibliographystyle{ACM-Reference-Format}
\balance
\bibliography{coride}

\end{document}